\documentclass[%
aip,
amsmath,amssymb,%
author-year,%
twocolumn%
]{revtex4}

\usepackage{graphicx}
\usepackage{dcolumn}
\usepackage{bm}

\usepackage[utf8]{inputenc}
\usepackage[T1]{fontenc}
\usepackage{newtxtext}
\usepackage[varvw]{newtxmath}

\usepackage{color}
\usepackage{bm}
\usepackage{amsmath}
\usepackage{amsfonts}
\usepackage{amssymb}
\usepackage{bm}
\usepackage{dsfont}
\usepackage{float}
\usepackage{upgreek}

\usepackage{titlesec}
\renewcommand{\vec}[1]{\ensuremath{\bm{#1}}}
\newcommand{\bvec}[1]{\ensuremath{\bm{#1}}}
\newcommand{\mat}[1]{\ensuremath{{\bm{#1}}}}

\newcommand{\parac}[1]{\ensuremath{#1_{z}}}
\newcommand{\perpc}[1]{\ensuremath{#1_{\perp}}}

\newcommand{\at}[2]{\ensuremath{\left (  #1 \right ) _{#2}}}

\newcommand{\dn}[1]{\ensuremath{\mathrm d^{#1}}}
\newcommand*{\noaddvspace}{\renewcommand*{\addvspace}[1]{}}
\addtocontents{lof}{\protect\noaddvspace}
\newcommand{\ub}{\ensuremath{\hat{\bvec z}}}
\newcommand{\ux}{\ensuremath{\hat{\bvec x}}}

\begin{document}
\preprint{AIP/123-QED}

\title{Energy balance for 6D kinetic ions with adiabatic electrons} 

\author{M. Raeth}
\email{Mario.Raeth@ipp.mpg.de}
\affiliation{Max Planck Institute for Plasma Physics, Boltzmannstr. 2, 85748 Garching, Germany}

\author{K. Hallatschek}
\affiliation{Max Planck Institute for Plasma Physics, Boltzmannstr. 2, 85748 Garching, Germany}

\date{\today}
\begin{abstract} 
This paper investigates the energy fluxes for the 6D kinetic Vlasov system. We
introduce a novel method for calculating particle and energy flows within this
framework which allows for the determination of energy and particle fluxes, as
well as the Poynting flux, directly from the system's moments such as kinetic
energy density, momentum transfer tensor. The fluxes computed using the new
method exhibit fewer gyrooscillations. This approach also enables the
identification of both the gyrokinetic $\vec{E} \times \vec{B}$ heat flux and
additional non-gyrokinetic contributions, while simultaneously reducing inherent
gyrooscillations in the energy and particle fluxes. Our semi-Lagrangian solver
for the 6D kinetic Vlasov system, features a highly efficient scheme to address
the $\vec v \times \vec B$ acceleration from the strong background magnetic
field allows for the simulation of plasma waves and turbulence with frequencies
extending beyond the cyclotron frequency, independent of gradient strength or
fluctuation levels. The solver has been rigorously tested in the low-frequency
regime for dispersion relations and energy fluxes in both linear and nonlinear
scenarios. 
\end{abstract}
\maketitle
\section{Introduction}
Kinetic models can describe phenomena within a tokamak plasma across different
scales however, until recently, we lacked the computational capacity for the
fully kinetic simulations. Gyrokinetic models used in codes like CGYRO
\cite{candy_high-accuracy_2016}, GENE \cite{jenko_massively_2000} and ORB5
\cite{lanti_orb5_2020}, agree well with experiments in fusion device cores but
face challenges in high-gradient regimes like the tokamak plasma edge. Most
models rely on a $\delta f$ approximations and break down in these regions of
steep gradients and high fluctuation amplitudes. In recent years, there is a
growing interest in the instability of ion Bernstein waves (IBWs)
\cite{yoon_bernstein_2014,noreen_ion_2019,rath_beyond_2023}, including the
excitation of turbulent IBWs in steep gradient regions such as the plasma edge
\cite{raeth_high_2023}. To evaluate the importance of these modes it is
important to understand the energy fluxes in the system. \\
This work provides a comprehensive analysis of the kinetic and electric energy
balance for 6D kinetic ions coupled with adiabatic electrons. We introduce a
novel method for computing the energy transport which reduces the amount of
Larmor oscillations and clearly shows the contributions of different moments
of the distribution function to the energy flux, including the
$\vec E \times \vec B $ energy flux familiar from gyrokinetics, thus
establishing a direct connection between the two models. In addition to the
analytical description, we compare the new method with the straight-forward
summation of the energy fluxes for ITG turbulence simulated with the 6D
kinetic Vlasov code BSL6D
\cite{schild_convergence_2024,schild_performance_2024}.\\
After a brief introduction to the physical system in section
\ref{sec_physical_system}, we describe the various energy transport channels,
beginning with the energy flux in adiabatic electrons and the Poynting flux
(section \ref{sec_energy_balance}), followed by a detailed description of
particle and energy transport for kinetic ions. Section
\ref{sec_numerical_results} presents simulation results that compare the newly
derived energy fluxes with the straight-forward summation.

\section{Physical system \label{sec_physical_system}} 
The Vlasov equation describes the motion of a plasma in presence of
electromagnetic fields. We consider ions in a constant and homogeneous magnetic
field $\vec B_0 = \hat{\vec z}$ represented by a 6D distribution function $f$,
with an electric field $\vec E$ originating from the interaction with adiabatic
electrons. The target of our 6D simulations is the collisionless ion kinetic
equation in dimensionless variables ($\rho_i= v_{\text{th}}=n_0=T=1$, with the
ion Larmor radius $\rho_i = \sqrt{m_i T}/(eB)$, the ion thermal velocity $v_{\text{th}} = \sqrt{T/m_i}$, the ion
background density $n_0$ and the ion background temperature $T$)
\begin{equation}
  \partial_t f+\vec v\cdot\nabla f+(-\nabla\phi+\vec v\times\vec  \hat{\vec z})
  \cdot\nabla_{\vec v} f=0.\label{eq_vlasov}
\end{equation}
For simplicity, we assume that the electrons are adiabatic $f_e = e^{\phi}f_M
\approx (1 + \phi) f_M$ [assuming a Maxwellian background distribution $f_M =
\exp(- {v^2}/{2})/\left(2\pi^{3/2}\right)$] and the plasma is quasineutral
$n_e=n$  (with the electron density $n_e$), the electrostatic field is given by
\begin{align}
	\phi =  n-1, \quad \quad \quad\vspace{2cm}  n = \int f \dn3v. \label{eq_field_equation}
\end{align}
\section{Energy balance \label{sec_energy_balance}}
\subsection{Electron energy balance}
Before computing the energy fluxes of the kinetic ions, we are focusing on the
contribution to the energy balance by the electrons. In the following, electron
quantities carry an index '$e$' ($f_e, n_e, T_e $, etc.), while quantities
without index ($f, n, T$,  etc.) refer to the ions.\\
For the entire derivation, we consider the quasineutral limit
($\epsilon_0\rightarrow 0$), the energy contained in the electric field
$\epsilon_{\text{\vec E}} = \epsilon_0 E^2/2$ and the Coulomb term $\epsilon_0
\Delta \phi $ are zero. The change of the electron energy density
$\partial_t{\epsilon}_e$  is sum of the divergence of the energy density flux
$\vec Q_e$ and the work performed on the electrons by the electric field $\vec E
\cdot \Gamma_e$
\begin{align}
	\partial_t  \epsilon_e = - \nabla \cdot \vec Q_e -\vec E \cdot \vec \Gamma_e. \label{eq_general_energy_balance}
\end{align}
When inserting $\nabla \phi = - \vec E$ in the last term of
\eqref{eq_general_energy_balance}, it can be rewritten as
\begin{align}
	 - \vec E \cdot \vec \Gamma_e =  (\nabla \phi) \cdot \vec \Gamma_e = \nabla \cdot (\phi \vec \Gamma_e) -  \phi \nabla \cdot \vec  \Gamma_e,
\end{align}
which can be further modified using the continuity equation 
\begin{align}
	\partial_t n = -\nabla \cdot \vec\Gamma_e \label{eq_continuity_equation}
\end{align}
resulting in
\begin{align}
	- \vec E \cdot \vec \Gamma_e =  \nabla \cdot (\phi \vec \Gamma_e) + \phi  \partial_t n_e.
\end{align}
With quasineutrality equation for adiabatic electrons $\phi = n_e - 1 =n-1$
(assuming $T_e =1$) results in 
\begin{align}
	\partial_t \frac{(n-1)^2}2 = -  \nabla \cdot (\phi \vec \Gamma_e)  - \vec E \cdot \vec \Gamma_e  \label{eq_energy_balance}
\end{align}
By comparing equations \eqref{eq_general_energy_balance} and
\eqref{eq_energy_balance}, we can identify expressions for the energy electron
density \cite{hallatschek_thermodynamic_2004}
\begin{align}
\epsilon_e = \frac{(n-1)^2}{2},
\end{align} 
and subsequently, the energy flux density is given by 
\begin{align}
	\vec Q_e = \phi \vec \Gamma_e. \label{eq_electron_energy_flux}
\end{align} 
(The adiabatic electrons can be interpreted as local capacitors, since $\phi \propto \delta n$, but for consistency with the Poynting flux, we need to account for the electron currents).
\eqref{eq_electron_energy_flux}
\subsection{Poynting flux}
In addition to the energy flux in the electrons, the Poynting flux $\vec S =
\vec E \times \vec B/{\mu_0}$ has to be considered. We rewrite the Poynting
flux using $\vec B = \ub + \delta B$ and $\vec E = -\nabla \phi + \partial_t
\vec A$, as
\begin{align}
\vec S &= \frac{1}{\mu_0} \vec E \times \vec B\notag\\ &= \frac{1}{\mu_0} (-\nabla \phi \times \vec B + \partial_t{\vec A} \times \vec B)\notag\\
&= \underbrace{-\frac{1}{\mu_0} \nabla \times (\phi \vec B) }_{O(\frac{1}{\mu_0})}+\underbrace{ \phi \frac{\nabla\times \vec B}{\mu_0} }_{O(1)}+ \underbrace{\frac{\partial_t{\vec A}}{\mu_0} \times \ub}_{O(1)} +\underbrace{\frac{\partial_t{\vec A} \times \delta \vec B}{\mu_0}}_{O(\mu_0)}. \label{eq_poynting_limit}
\end{align}
Since $\Delta \vec A = \mu_0 \vec j$ and $\nabla \times \vec B = \mu_0 \vec j$, we
know that $\delta \vec B\ = O(\mu_0)$ and $\vec A = O(\mu_0)$ and thus,
$\partial_t{\vec A} \times \delta \vec B/{\mu_0} = O(\mu_0)$. In the
electrostatic limit, $\mu_0\rightarrow 0$, most terms of \eqref{eq_poynting_limit}
vanish. However, the first term $\nabla\times (\phi \ub)/{\mu_0}\rightarrow
\infty$ because the background magnetic field is externally imposed rather than
generated by plasma currents. Since the resulting term is
divergence free, it does not
contribute to the energy transport and can be ignored in the following
discussion of turbulent transport balance. Using Amp\`eres law, where the displacement current vanishes because of the quasineutral limit, the Poynting flux becomes 
\begin{align}
	\vec S = - \nabla \times \frac{\phi \vec B}{\mu_0} + \phi \vec j + \frac{\partial_t {\vec A}}{\mu_0}\times \ub, 
\end{align}	
where $\vec j = \vec \Gamma - \vec \Gamma_e$. The last term
$\partial_t{\vec A} \times \ub /\mu_0$ can also be omitted from the turbulent
transport balance as we are interested in quasi steady state energy fluxes,
and the total time derivatives vanishes upon time averaging. The term
$\phi \vec j$ seems problematic, as adding a constant to the electrostatic
potential would change the Poynting flux. However, its divergence is invariant
when adding a constant $\chi$ to $\phi$,
\begin{align}
\nabla \cdot \vec S  = \nabla \cdot  (\phi +  \chi)\vec j = \nabla \cdot (\phi \vec j ) + \chi \nabla \cdot \vec j = \nabla \cdot (\phi \vec j ),
\end{align}
since quasineutrality implies $\nabla \cdot \vec j = 0$. 
\section{Ion transport equations \label{sec_transport_equations}}
To arrive at the various expressions for the fluxes, we compute the time
derivative of the first four moments of the distribution function $f$ using the
Vlasov equation \eqref{eq_vlasov}.
\subsection{Particle density}
The zeroth moment of the Vlasov equation yields the continuity equation \eqref{eq_continuity_equation} with the particle flux $\vec \Gamma = \int \vec v f
\dn 3v$, written in fluid quantities is $\vec \Gamma = n \vec u$ (with the mean
velocity $\vec u$).\\
The first moments of the Vlasov equation can be rewritten as
\begin{align}
\partial_t {\vec \Gamma} &= \int \vec v \partial_t f  \dn 3 v\notag\\ &= -\int \vec v \vec v \cdot \nabla f \dn 3 v-\int \vec v (-\nabla \phi + \vec v \times \hat{\vec z})\cdot \nabla_v f \dn 3v.
\end{align}
The first integral results in the divergence of the momentum transfer tensor $\nabla \cdot
\mat {\Pi} = \nabla \cdot \int \vec v \vec v f \dn 3 v$, which can be expressed
in terms of fluid quantities $\mat \Pi = \mat \pi + \mathds 1 p + \vec u \vec u
n $  (with the stress tensor $\mat \pi$, pressure $p$ and mean velocity $\vec
u$).  The Lorentz force term is modified using integration by parts, resulting
in 
\begin{align}
	\partial_t {\vec \Gamma} &=-\nabla \cdot  \mat \Pi - n \nabla \phi  + \vec \Gamma \times \ub.
\end{align}
After computing the cross product with the magnetic field $\ub$, the Grassmann
identity can be applied to obtain an expression for the perpendicular particle
flux
\begin{align}
	\perpc {\vec \Gamma} = -\partial_t {\vec \Gamma} \times \ub - \nabla \cdot \mat \Pi \times \ub - (\nabla \phi \times \ub) n. \label{eq_particla_flux}
\end{align}
The last term is the $\vec E \times \vec B$ particle flux known from
gyrokinetics. Since we are interested in quasi steady-state particle fluxes, a
time average of the flux is considered, eliminating the time derivative. The
other terms vanish when a flux surface average, defined as $\langle \cdot
\rangle = \int \cdot , \mathrm{d}y , \mathrm{d}z/({L_y L_z} )$, is applied the
expression can be written as
\begin{align}
\langle  \nabla \cdot \mat \Pi \times \ub  \rangle + \langle \nabla \phi \times \ub  n
\rangle  = & \ux (\left \langle n \partial_y  \phi \right \rangle + \partial_x\langle \Pi_{xy} \rangle) \notag\\-& \hat{\vec y}(\left \langle n \partial_x  \phi \right \rangle + \partial_x \langle \Pi_{xx}\rangle) .
\end{align}
When the adiabatic assumption \eqref{eq_field_equation} is inserted, we obtain
\begin{align}
	\langle \perpc {\vec \Gamma}\rangle = & -\ux  \partial_x\langle \Pi_{xy} \rangle+ \hat{\vec y} \partial_x\left(\left \langle   \frac{n^2}2 \right \rangle + \langle \Pi_{xx}\rangle\right) .
\end{align}
In flux tube scenarios, both terms vanish, as they are total derivatives with
respect to $x$ and typically globally averaged quantities are considered.
\subsection{Energy density}
The total energy density of the system is given by the sum of the kinetic energy
of ions $\epsilon = \int {v^2}f/2 \dn3v$ (expressed in thermodynamic quantities
$\epsilon = (3/2) p + {u^2}n/2 $ with $p=nT$) and the electron energy
$\epsilon_e = {(n-1)^2}/2$ derived in section \ref{sec_energy_balance}. The time
derivative of the ion kinetic energy is again computed using the Vlasov equation
(\ref{eq_vlasov})
\begin{align}
  \partial_t \epsilon =& \int\frac {v^2}2 \partial_t f  \mathrm{d}^3v  \notag \\
  =& - \int \frac {v^2}2 \vec v \cdot \nabla f  \mathrm{d}^3v\notag \\
  & -  \int\frac {v^2}2 (-\nabla \phi +\vec v \times \ub)\cdot \nabla_{\vec v} f  \mathrm{d}^3v.
\end{align}
Integrating the second integral by parts, yields
\begin{align}
  \partial_t \epsilon &= - \nabla \cdot \vec Q -  \int \vec v\cdot ( \nabla \phi) f  \mathrm{d}^3v, \label{eq_ion_energy_density}
\end{align}
with the total local ion energy flux density $\vec Q = \int \vec v  {v^2}/{2}f
\dn 3v$. Similar to the momentum transfer tensor, the total ion energy flux can
be written in terms of the macroscopic fluid quantities $	\vec Q 	= \vec q +
{u^2}n\vec u/{2}   + \vec u \cdot \mat \pi  +(5/2)p \vec u $, with the heat flux
$\vec q  = \int (\vec v - \vec u) (\vec v -\vec u)\cdot (\vec v -\vec u)/2 f \dn
3 v$. The second term in \eqref{eq_ion_energy_density} is modified to be
expressed as a divergence of the entire integral
\begin{align}
  \partial_t \epsilon &= -\nabla \cdot \vec Q - \nabla  \cdot( \phi \vec \Gamma) + \phi \nabla \cdot \vec \Gamma,
\end{align}
and inserting \eqref{eq_continuity_equation} results in
\begin{align}
	\partial_t (\epsilon + \phi  \partial_t n) &=  -\nabla \cdot \vec Q - \nabla \cdot (\phi \vec \Gamma). \label{eq_2nd_moment_vlasov}
\end{align}
Utilizing the field equation (\ref{eq_field_equation}), the electron energy
density can be inserted $\partial_t\epsilon_e = \partial_t(n-1)^2/2 = \phi
\partial_t n$. Apart from the local energy flux density $ \vec Q$, an additional
energy flux $ \phi \vec \Gamma$ appears which, can be identified as
the sum of the electron energy flux and the Poynting flux
$\vec S + \vec Q_e =  \phi \vec \Gamma$. \\
The energy flux in the gyrokinetic system is described by the $\vec E \times
\vec B$ heat flux. This result should be recovered for gyrokinetic modes in the
6D kinetic system. Thus, the kinetic energy flux of the ions should contain the
$\vec E \times \vec B$ heat flux plus an additional contribution, which cancels
$\phi \vec \Gamma$ under the correct conditions.\\
To establish the connection of the energy flux $\vec Q$ and the gyrokinetic
$\vec E \times \vec B$ heat flux, the energy flux can be split into its
contributions. For this purpose, the time derivative $\partial_t {\vec Q}$ is
computed
\begin{align}
	\partial_t {\vec Q} &= \int\vec v\frac{v^2}{2}  \partial_t f  \dn 3 v\notag\\ &= -\int\frac{v^2}{2}  \vec v \vec v \cdot \nabla f \dn 3 v-\int \frac{v^2}{2} \vec v (-\nabla \phi + \vec v \times \hat{\vec z})\cdot \nabla_v f \dn 3v \notag \\
	&= - \nabla \cdot \mat \Pi^*-\int \frac{v^2}{2} \vec v (-\nabla \phi + \vec v \times \hat{\vec z})\cdot \nabla_v f \dn 3v,
	\end{align}
with $\mat \Pi^* = \int {v^2}\vec v \vec v f/{2} \dn3v$. The Lorentz force term
is modified using integration by parts and inserting the expression for kinetic
energy density $\epsilon$, and the momentum transfer tensor $\mat \Pi$
\begin{align}
  -\int& \frac{v^2}{2} \vec v\left[ (-\nabla \phi + \vec v \times \ub) \cdot \nabla_{\vec v} f \right] \dn 3 v \notag \\
  =& -\int \left(\nabla \phi \cdot \vec v\right)  \vec v f \dn 3v - \int \frac{v^2}2 \nabla \phi f \dn 3 v + \int (\vec v\times \ub) \frac{v^2}{2} f\dn 3 v \notag \\ 
  =& -\mat \Pi \cdot \nabla \phi - \epsilon \nabla \phi + \vec Q \times \ub.
\end{align}
After computing the cross product with the magnetic field $\partial_t \vec Q
\times \hat{\vec z}$, the Grassmann identity can be applied to obtain an
expression for the perpendicular energy flux
\begin{align}
	\perpc {\vec Q}  = -&\partial_t{ \vec Q}\times \hat{\vec z}- \nabla \cdot \mat \Pi^*\times \hat{\vec z}\notag \\&-  \nabla \phi \cdot\mat \Pi  \times \hat{\vec z} - \nabla \phi\times \hat{\vec z} \epsilon. \label{eq_energy_flux_contributions}
\end{align}
Computing the time and flux surface average of this equation yields 
\begin{align}
	\langle \vec Q_\perp \rangle = -\langle  \nabla \phi\times \hat{\vec z} \epsilon \rangle - \langle\nabla \phi \cdot\mat \Pi  \times \hat{\vec z} \rangle  - \langle \nabla \cdot \mat \Pi^*\times \hat{\vec z} \rangle. \label{eq_total_energy_flux}
\end{align}
Next to the $\vec E \times \vec B$ heat flux $\vec Q ^{\vec E \times \vec B} =
\nabla \phi\times \hat{\vec z} \epsilon $, two more contributions can be
identified. In order to compare their relevance for the total energy balance
\eqref{eq_2nd_moment_vlasov}, we also express the Poynting flux in terms of the
contributions from equation \eqref{eq_particla_flux}
\begin{align}
	\perpc{\vec S} = \phi\perpc {\vec \Gamma} =- \phi\partial_t {\vec \Gamma} \times \ub - \phi\nabla \cdot \mat \Pi \times \ub - \phi (\nabla \phi \times \ub) n. 
\end{align}
To study quasi steady-state energy flux, the quantities are time-averaged,
effectively removing the time derivative of the energy flux, $\partial_t{\vec
Q}$. In contrast, the time average of $\phi\partial_t{\vec \Gamma}$ does not
vanish. In summary, the flux surface average of the total energy balance from
equation \eqref{eq_2nd_moment_vlasov} is given by
\begin{align}
	\partial_t \langle \epsilon + \phi \partial_t n\rangle =&  -\ux \partial_x\cdot \langle \vec Q + \phi \vec \Gamma\rangle\notag \\
	= -\ux \partial_x\cdot \langle& {-\nabla \phi\times  \hat{\vec z}   \epsilon}   -\nabla \phi \cdot\mat \Pi  \times \hat{\vec z} - \nabla \cdot \mat \Pi^* \times \ub \notag \\&   
	- \phi n\nabla \phi \times \ub- \phi\nabla \cdot \mat \Pi \times \ub -\phi\partial_t {\vec \Gamma} \times \ub\rangle 
\end{align}
The two terms containing the momentum flux tensor $\mat \Pi$ can be combined
\begin{align}
	\langle( \mat \Pi \cdot\nabla \phi) \times \hat{\vec z} \rangle   + \langle (\phi\nabla \cdot \mat \Pi) \times \ub \rangle =  \langle \nabla \cdot (\phi \mat \Pi) \rangle \times \hat{\vec z} 
\end{align}
resulting in total energy balance 
\begin{align}
		\partial_t &\langle \epsilon + \phi \partial_t n\rangle\notag \\&=  - \ux \partial_x \cdot\left \langle {-\epsilon\nabla \phi}   -   \nabla \cdot (\phi \mat \Pi + \mat \Pi^*)  -  \phi \partial_t{\vec \Gamma} - \phi n \nabla \phi  \right\rangle \times \ub. \label{eq_total_energy_balance_fs}
\end{align}
The last term $\langle \phi n(\nabla \phi \times \ub)\rangle$ is the Poynting
flux caused by the $\vec E \times \vec B$ particle flux which averages away in
the flux surface average for adiabatic electrons
\begin{align}
	\langle \phi (\nabla \phi \times \ub) n \rangle = \langle n^2 \nabla \phi \rangle \times \ub =  \nabla\langle \frac{n^3}3 \rangle \times \ub =0,
\end{align}
when the field equation \eqref{eq_field_equation} is inserted, because odd
powers of a perturbed quantity average out. Similarly to the particle flux
induced by the momentum transfer tensor, $\langle \nabla \cdot (\phi \mat \Pi +
\mat \Pi^*) \rangle \times \hat{\vec z}$ can be further simplified,
\begin{align}
	\partial_x \ux \cdot \langle \nabla \cdot (\phi \mat \Pi + \mat \Pi^*) \rangle \times \hat{\vec z} = \partial_x^2 \langle \phi \Pi_{xy} + \Pi_{xy}^*\rangle.
\end{align}
Inserting these two simplification into \eqref{eq_total_energy_balance_fs}
results in the energy balance for kinetic ions with adiabatic electrons
\begin{align}
\partial_t \left \langle \epsilon + \epsilon_e \right\rangle = -\partial_x\left \langle - \epsilon \partial_y \phi  -  \partial_x ( \phi \Pi_{xy} + \Pi_{xy}^*) - \phi \partial_t \Gamma_y\right \rangle.
\end{align}
In summary, the total energy flux in the fully kinetic systems contains two
contributions beyond the well-known gyrokinetic heat flux. \\
Firstly, the additional energy flux induced by $\mat \Pi$ and $\mat \Pi^*$. This
term arises from the radial dependency of the background profiles. In global
gyrokinetic simulations, a similar energy flux would occur, driven by the $\Pi$
and $\Pi^*$ tensors resulting from the gyrotransformation from gyrocenter to
particle coordinates.\\
For $k_x = 0$, this term vanishes, leaving only the kinetic energy flux
associated with the $\vec{E} \times \vec{B}$ heat flux. The remaining difference
compared to gyrokinetics is the Poynting flux component $\phi \partial_t{\vec
\Gamma}$. This can be neglected for gyrokinetic modes with $\omega \ll 1$, but
it might become relevant for high-frequency modes with $\omega \gtrapprox 1$,
such as IBW turbulence \cite{raeth_high_2023}.
\section{Simulation of energy fluxes \label{sec_numerical_results}}
\subsection{Energy flux for gradients in Boussinesq approximation}
\begin{figure}
	\includegraphics[width = 0.45\textwidth]{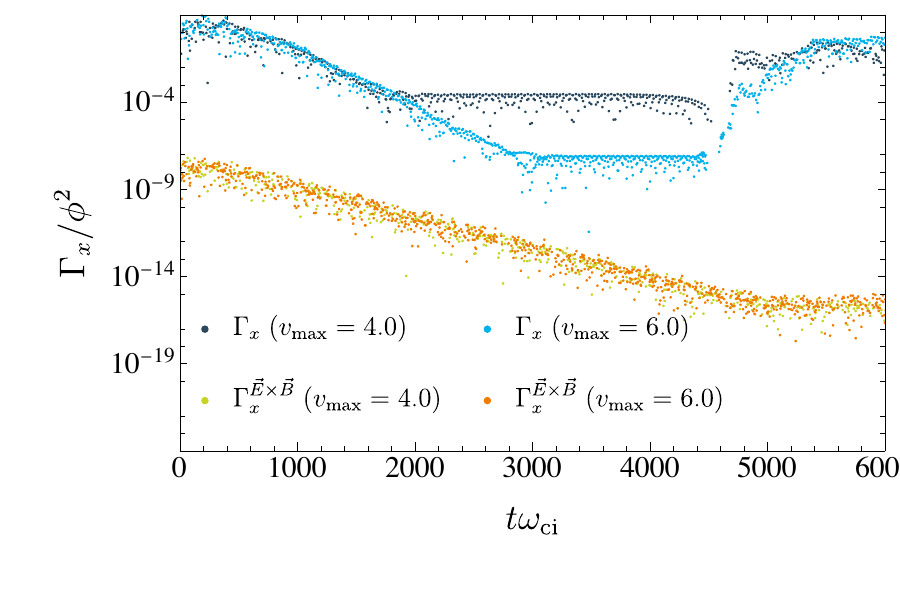}
	\caption{Comparison of flux surface averaged particle flux with $\vec E
	\times \vec B$ particle flux for simulation with $v_{\text{max}}=4$ and
	$v_{\text{max}}=6$. All quantities are normalized to the square amplitude of
	the electrostatic potential \label{fig_particle_flux}}
\end{figure}

\begin{figure}
	\includegraphics[width = 0.45\textwidth]{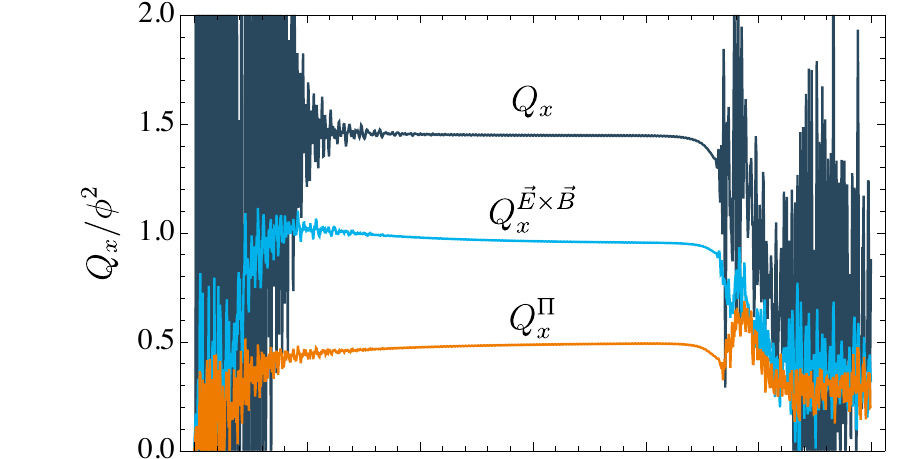}
	\includegraphics[width = 0.45\textwidth]{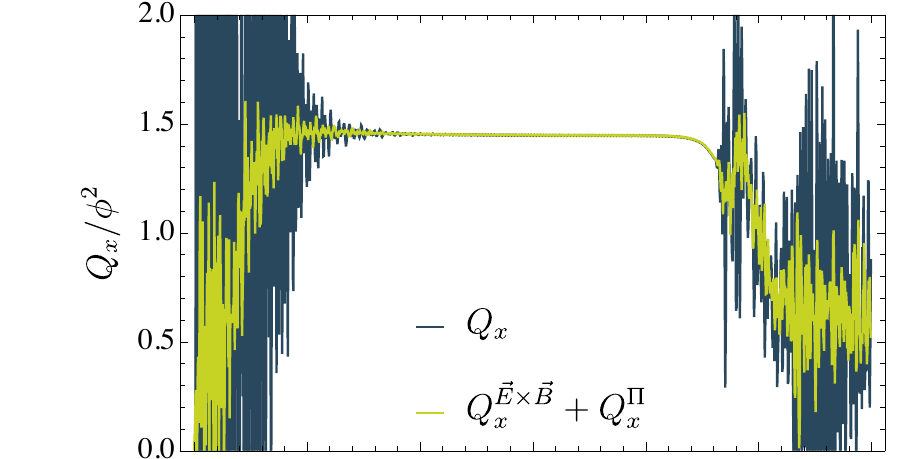}
	\includegraphics[width = 0.45\textwidth]{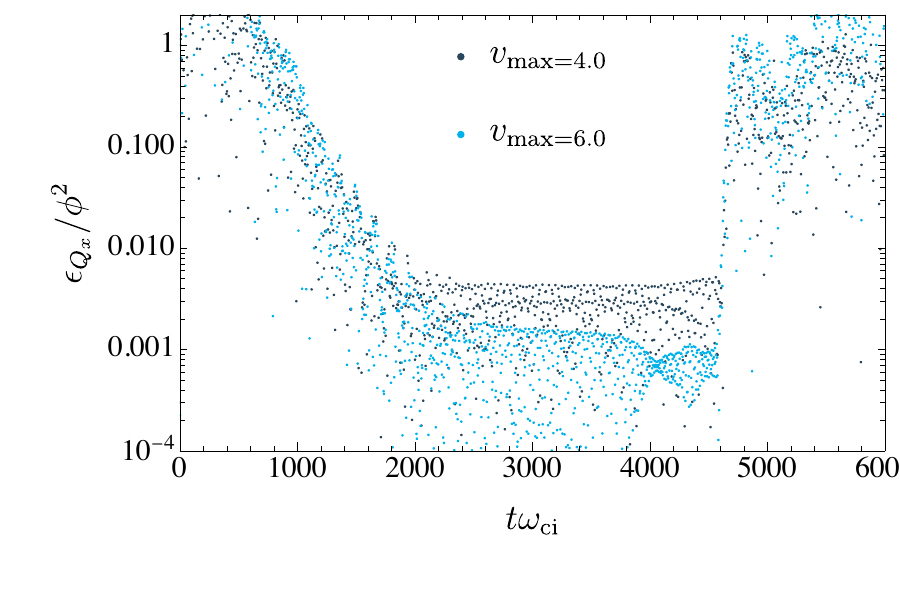}
	\caption{Comparison of the energy flux $\vec Q=\int {v^2} f/2 \dn 3v$ with
	its contributions derived in equation \eqref{eq_energy_flux_contributions}
	(top, middle) and error for simulation with $v_{\text{max}}=4$ and
	$v_{\text{max}}=6$ (bottom). All quantities are normalized to the square
	amplitude of the electrostatic potential
	\label{fig_energy_fluxes_boussinesq}}
\end{figure}

For the first numerical test, we investigate the ion temperature gradients (ITG)
instability previously shown in \cite{raeth_simulation_2024}.  For the
simulation, we introduce a temperature gradient source term on the RHS
\begin{align}
	\partial_t \delta f + \vec v\cdot \nabla \delta f + \left[-\nabla \phi+(\vec v\times \hat{\vec z})\right] \cdot \nabla_v \delta f =  \nabla \phi\cdot \nabla_v g_0,
\end{align}
where $g_0$ is the background distribution function 
\begin{align}
	g_0(\vec R, \perpc v, \parac v) =\left( \frac{n(\vec R)}{2\pi T(\vec R)}\right)^{\frac 32} \exp\left(-\frac{ [\perpc v ^2 +
		\parac v^2]}{2 T(\vec R)}\right) \label{eq_maxwellian_profiles}
\end{align}
and $\vec R$ is the gyrocenter coordinates \( \vec{R} = \vec{r} - \vec{\rho} \)
(where \( \vec{\rho} \) is the Larmor radius vector). The gradient on the
right-hand side is given by (for clarification, we introduce a notation to
specify which variable is held constant for the partial derivative, i.e.
$\at{\nabla_{\vec v} f}{\vec r} $ for a derivative with respect to $\vec v$ at
fixed $\vec r$)
\begin{align}
	\at{\nabla_v g_0}{\vec r} &= \ub \times \at{\nabla_{\vec R} g_0}{\vec v} + \at{\nabla_v g_0}{\vec R}  =: {\vec v^*}g_0 + \at{\nabla_v g_0}{\vec R}\notag \\
	&= -g_0 \ub \times \left( \frac{ \nabla T}{T} \frac{3 -  (\perpc v ^2 + \parac v^2)}{2 }\right) +  \at{\nabla_v g_0}{\vec R} .\label{eq_source_term}
\end{align}
The limit that the gradient length is infinite $T(\vec R) \rightarrow T = 1$ is
taken, such that the background distribution is independent of $\vec R$. Thus,
$g_0$ is replaced with the homogeneous Maxwellian distribution $f_M = \exp(-
{v^2}/{2})/\left(2\pi^{3/2}\right)$. \\
To numerically measure the transport values, we have conducted a 2D3V simulation
of an ITG instability. The box has been chosen with dimensions \(  L_y = 20/3\pi
\), $L_z = 240\pi $ and \( (N_y,N_z) = (64,16) \). It has been shown in previous
works that due to the small wave numbers in parallel direction, the small number
of points in $z$-direction is sufficient \cite{rath_beyond_2023}. For the given
box dimensions, the smallest perpendicular wave number is \(k_{y,0} = 0.3 \).
The length in the parallel direction is determined by the wave number of the
fastest-growing mode in the system, which depends on the temperature gradient.
Employing a temperature gradient of \( \partial_x \log T= 0.1 \), we obtained an
ideal parallel wave number of \( k_{z} = {1}/{120} \).\\
Velocity space is symmetric in all directions, with \( v_{\mathrm{max}} = 4 \) (or  \( v_{\mathrm{max}} = 6 \)) and \( (N_{v_x},N_{v_y},N_{v_z}) = (33,33,33)
\). We applied periodic boundary conditions across all six dimensions. The
interpolation utilizes a 7th-order Lagrange stencil for the velocity directions
and an 8th order stencil for the spatial directions. 
\begin{figure}
	\includegraphics[width = 0.45\textwidth]{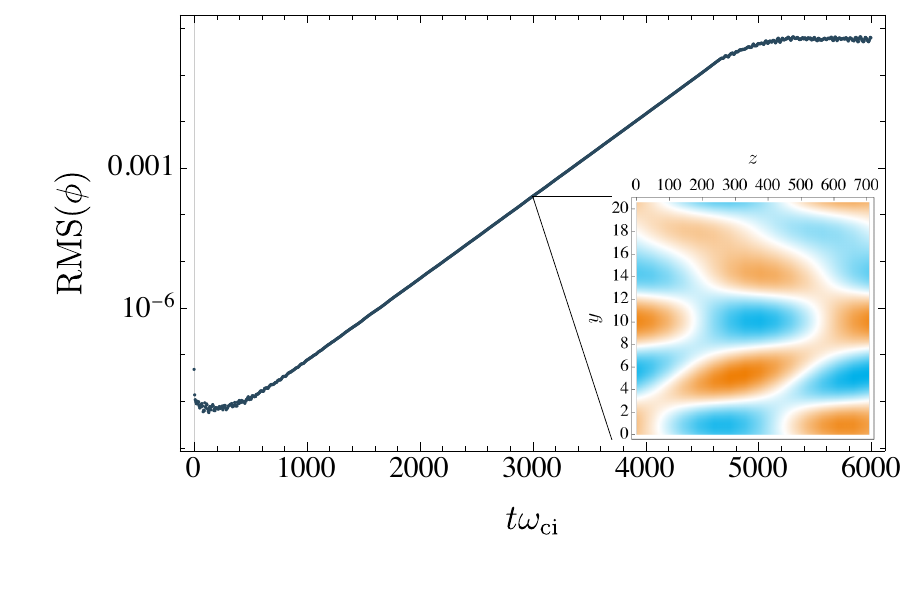}
	\caption{Time evolution of electrostatic potential amplitude with snapshot
	of perturbation in $y-z-$ plane  \label{fig_amplitude_phi}}
\end{figure}
From the simulation, we can extract the necessary moments of the distribution
function and compute the relevant energy fluxes for our comparison. Figure
\ref{fig_amplitude_phi} shows the root-mean-square (RMS) of the electrostatic
potential as a function of time. We see a linear growth for the time between
$t\sim 1000 - 4500$ and nonlinear saturation at around $t\sim 5000$. As an inset
we show a snapshot at $t=3000$ of the electrostatic potential in the $y-z-$plane,
showing the clear ITG mode structure, of the dominate mode with $\vec k =
(0,3/5, 1/120)$. \\
It has been discussed in section \ref{sec_transport_equations} that the flux
surface averaged particle flux is zero. This should be investigated first. For
this purpose we have calculated the particle flux $\langle\Gamma_x\rangle = \langle\int v_x f
\dn 3v\rangle$ and the $\vec E \times \vec B$ particle flux $\langle\Gamma_x^{\vec E
\times \vec B}\rangle= \langle \nabla \phi \times \ub n \rangle$. Figure
\ref{fig_particle_flux} shows both quantities normalized to the square of the
fluctuation amplitude of the electrostatic potential for two separate
simulations with different maximal velocities.\\
The results show that the particle flux is zero up to numerical errors, which
are constant (ratio with fluctuation amplitude decreases with growing
electrostatic potential). An error proportional to the square of the
electrostatic potential occurs at the later stage of the simulation. By
increasing the size of the velocity grid from $v_{\text{max}}=4$ to
$v_{\text{max}}=6$, this error can be reduced by 4 orders of magnitude,
suggesting that the error origins from the boundaries in velocity
space. 
\\In previous work, we have analytically calculated the $\vec E \times \vec B$
energy flux and the Poynting flux \cite{raeth_simulation_2024}. Here, we
investigate the equivalence presented in equation
\eqref{eq_energy_flux_contributions}. The top plot in figure
\ref{fig_energy_fluxes_boussinesq} illustrates the contributions from the $\vec
E \times \vec B = $ heat flux $Q^{\vec E \times \vec B} = \epsilon \nabla \phi \times \ub$, the stress-induced energy
flux $Q^{\mat \Pi} = \nabla \cdot \mat \Pi \times \ub$, and $\vec Q = \int \vec v {v^2}f/2 \dn 3v$. The middle and
lower plots compare the LHS and RHS of equation \eqref{eq_total_energy_flux} and
show the error for different $v_{\text{max}}$ values. All quantities are
normalized to the square of the perturbation amplitude of the electrostatic
potential.
The results highlight one of the main advantages of the presented method for
calculating energy fluxes. The directly computed energy flux $\vec Q  = \int
\vec v{v^2} f/2 \dn 3 v$ exhibits significantly more oscillations than the
computed $\vec E \times \vec B$ heat flux and the stress-induced energy flux.
These strong oscillations in the total flux account for the majority of the
error during the early initialization and the nonlinear stage of the simulation.
In the linear phase, however, the two calculation methods show very good
agreement. As for the particle flux, we observe an error arising from the
boundary in velocity space, which is proportional to the square of the wave
amplitude.
\subsection{Nonlinear gradients}
\begin{figure}
	\includegraphics[width = 0.45\textwidth]{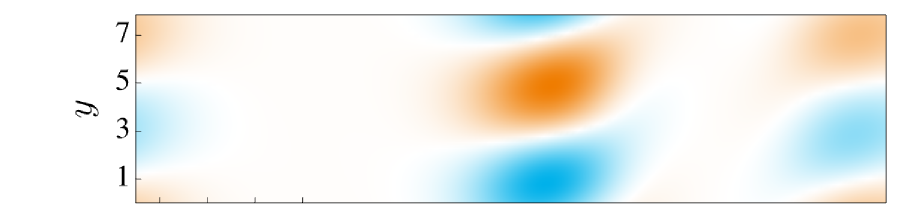}

	\includegraphics[width = 0.45\textwidth]{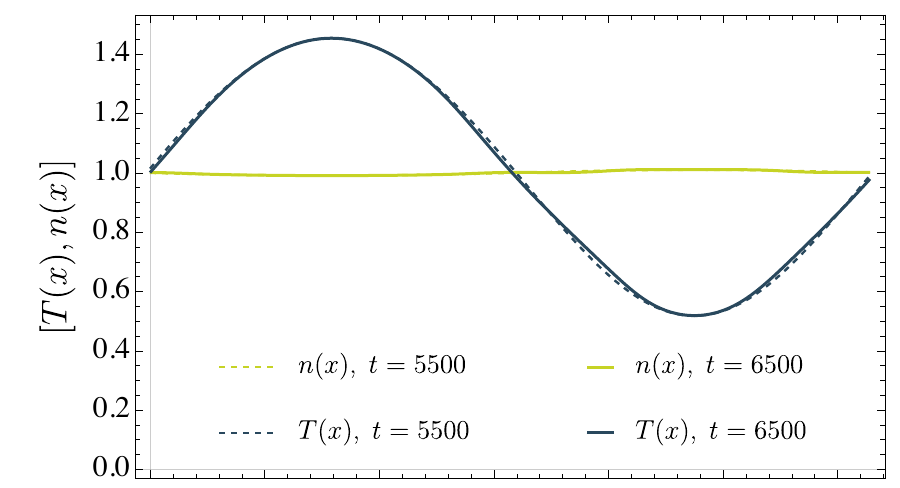}

	\includegraphics[width = 0.45\textwidth]{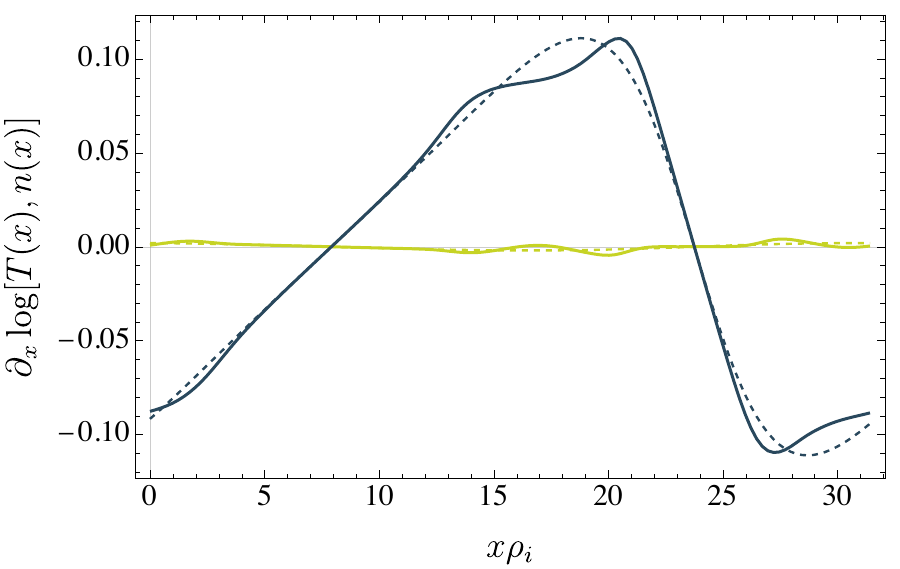}
	
	\caption{Cross section of electrostatic potential perturbation in
	$x-y-$plane in linear phase ($t=6000$) (top). Profiles (middle) and
	logarithmic gradient (bottom) of particle density and energy density for two
	different times \label{fig_profiles_decay}}
\end{figure}

\begin{figure}
	\includegraphics[width = 0.45\textwidth]{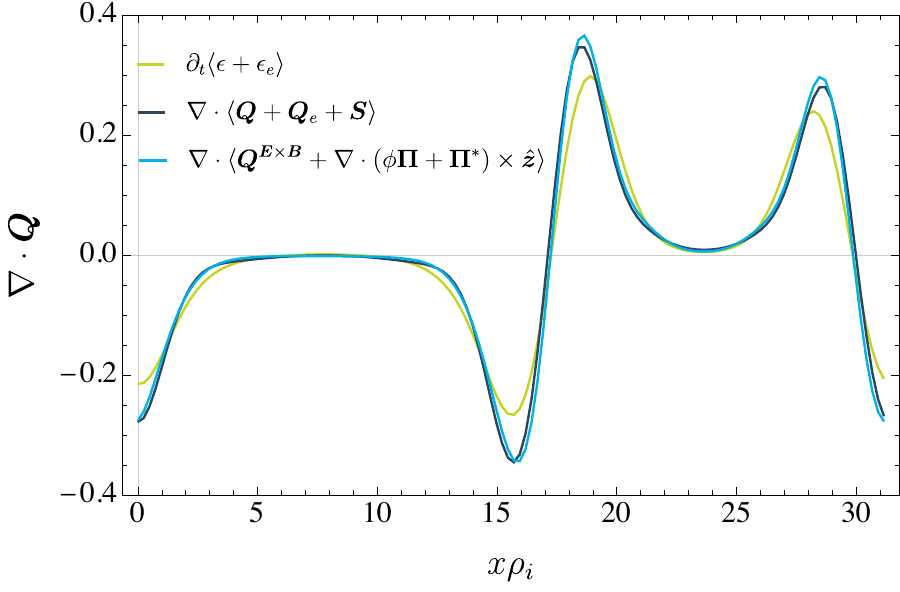}
	\caption{Comparison between change of energy density $\partial_t \langle
	E\rangle$ and divergence of energy fluxes $\nabla \cdot \vec Q$ at $t=6000$
	\label{fig_change_energy_density}}
\end{figure}

This section shifts away from the Boussinesq gradients described in the previous
section towards a nonlinear treatment of gradients in our simulation code. The
initial condition in the simulation is chosen such that the distribution
function contains density and temperature profiles as written in \eqref{eq_maxwellian_profiles}. The background profiles \( n(\vec R) \) and \( T(\vec R) \) are defined in gyrocenter coordinates \( \vec{R} = \vec{r} -
\vec{\rho} \) to remove oscillations with the Larmor frequency. To simplify the
treatment of boundary conditions, the profiles are periodic using sine-profiles
in the \( x \)-direction
\begin{align}
    n(x, v_y) = 1 + \kappa_n \sin\left(k_0 \left(x -  v_y\right)\right), \\
    T(x, v_y) = 1 + \kappa_T \sin\left(k_0 \left(x -  v_y\right)\right) \label{eq. temperature_profile}.
\end{align}
The initial background density gradient should not induce an electric field.
Therefore, modes with wave numbers parallel to the gradient are eliminated from
the electrostatic potential by subtracting the flux surface average
\begin{align}
    \phi = n - \langle n_0\rangle,
\end{align}
where $n_0$ is the density perturbation at $t=0$. When selecting parameters for
the background profile, it is crucial to ensure that the wavelength of the
unstable mode, \( \lambda_{\text{ITG}} = {2\pi}/{k_y} \), is shorter than the
typical gradient length of the profile, \( 1/L_T \sim {\partial_x \ln T(x)}
\).\\
For our initial simulation, we chose parameters: \( \kappa_n = 0 \), \( \kappa_T
= 0.5 \), and \( k_0 = 0.2 \), resulting in a temperature gradient \( \max_{x\in
[0,L_x]}\partial_x \log T(x) = 0.115 \), where \( L_x \) denotes the box length
in the \( x \)-direction. The simulation has been conducted on a box with a
length of $L=10\pi \times 2.5 \pi \times 240\pi$,
$N=128\times32\times16\times32\times32\times16$ and $\delta t = 0.02$.\\
The top plot in figure \ref{fig_profiles_decay} shows an $x-y-$cross-section of
the electrostatic potential just before the non-linear saturation, highlighting
two modes growing on the two slopes of the temperature profile. The second plot
shows density and temperature profiles at two different times, indicating a
slight decay. The decay is more evident when plotting the logarithmic gradients
of the profiles (Figure \ref{fig_profiles_decay}, bottom), where the erosion of
the gradients by the mode is visible. 
\\
The change of the local energy density can be compared to the divergence to the
energy fluxes computed from the different methods discussed in section
\ref{sec_transport_equations}. Figure \ref{fig_change_energy_density} shows a
comparison between the change of the total energy $\partial_t (\epsilon +
\epsilon_e)$ with the divergence total energy transport $\nabla \cdot (\vec Q +
\vec Q_e +  \vec S)$ directly computed from the distribution function and the
expression derived in equation \eqref{eq_total_energy_balance_fs} for $t=6000$.
At this point, the nonlinear saturation has begun, but the turbulence has not
fully developed yet (see figures \ref{fig_profiles_decay}). The plot shows a
very good agreement between the two different approaches for computing the
energy flux, showing that the decomposition of the energy fluxes in its
contributions in equation \eqref{eq_total_energy_balance_fs} is correct. 
Comparing the divergence of the energy flux to the change in local energy
density, we find that the local change rate of the total energy is lower by
approximately $10-15\%$. The modes are confined to a relatively narrow radial
area. The high wavenumber $k_x$ amplifies the effect of the radial dependency of
the heat flux, which could lead to a discrepancy. Furthermore, the strong radial
dependency of the heat flux can introduce numerical errors, owing to the limited
accuracy of Lagrange interpolation at high wavenumbers. The time derivative of
the changing total energy is computed using a finite difference derivative
averaged over exactly one Larmor period to remove the contributions from the
$\partial_t \vec Q \times \ub$ term in equation
\eqref{eq_energy_flux_contributions}. Increasing numerical resolution is
currently not feasible due to the high computational cost. 
\section{Summary}
In this work, we have introduced a novel method for calculating energy flux in
the 6D kinetic description of ions in a plasma. The method rewrites the energy
flux (kinetic energy flux and Poynting flux) in terms of other moments of the
distribution function. Figure \ref{fig_energy_fluxes_boussinesq} shows that this
approach reduces oscillations in the diagnostic, leading to more reliable
results. Additionally, it enables us to identify various contributions to the
energy balance and attribute physical interpretations to individual components.
This has helped to recognize the connection to the $\vec E \times \vec B$ heat
flux known from gyrokinetics and the contributions from the radial dependency of
background profiles. Moreover, it has enabled us to identify a potential
candidate for a difference in the energy flux for non-gyrokinetic high-frequency
modes. The effect of the $\phi \partial_t{\vec \Gamma}$ contribution, which has
been neglected due to the low frequency of ITG modes, requires further
investigation for waves with $\omega \gtrsim 1$.\\
The proposed framework could be extended to rewrite higher moments of the
distribution function, such as the momentum transfer tensor $\mat \Pi$ and the
$\mat \Pi^*$ tensor, using the same method. However, this would introduce a
dependency of the energy balance on higher-rank tensors, making the framework
less practical, and it would complicate the interpretation of the additional
terms. The conditions under which this continued expansion would converge remain
an open question.

\begin{acknowledgments}
This work has been carried out partly within the framework of the EUROfusion
Consortium, funded by the European Union via the Euratom Research and Training
Programme (Grant Agreement No 101052200 – EUROfusion). Support has also been
received by the EUROfusion High Performance Computer (Marconi-Fusion). Views and
opinions expressed are however those of the author(s) only and do not
necessarily reflect those of the European Union or the European Commission.
Neither the European Union nor the European Commission can be held responsible
for them.  Numerical simulations were performed at the MARCONI-Fusion
supercomputer at CINECA, Italy and at the HPC system at the Max Planck Computing
and Data Facility (MPCDF), Germany.
\end{acknowledgments}

\bibliographystyle{abbrv}
 \bibliography{Transport_Paper.bib}{}

\end{document}